\documentclass[aps,pra,twocolumn,superscriptaddress,groupedaddress,citeautoscript,preprintnumbers,floatfix]{revtex4}

\usepackage[utf8]{inputenc}
\usepackage[english]{babel}
\usepackage{amsfonts}
\usepackage{amsmath}
\usepackage[T1]{fontenc}
\usepackage{graphicx}
\usepackage[export]{adjustbox}
\usepackage{dcolumn}
\usepackage{bm}
\usepackage{tikz}
\usepackage{hyperref} 
\usepackage{braket}
\hypersetup{
  colorlinks = true,
  urlcolor = blue,
  linkcolor = blue,
  citecolor = blue
}

\usepackage{soul}
\usepackage[normalem]{ulem}

\begin{document}
\title{Tensor Network based Gene Regulatory Network Inference for Single-Cell Transcriptomic Data}

\author{Olatz Sanz Larrarte}
\email{osanzl@unav.es}
\affiliation{Department of Basic Sciences, Tecnun - University of Navarra, 20018 San Sebastian, Spain.}

\author{Borja Aizpurua}
\affiliation{Department of Basic Sciences, Tecnun - University of Navarra, 20018 San Sebastian, Spain.}
\affiliation{Multiverse Computing, Paseo de Miramon 170, E-20014 San Sebastian, Spain.}

\author{Reza Dastbasteh}
\affiliation{Department of Basic Sciences, Tecnun - University of Navarra, 20018 San Sebastian, Spain.}

\author{Ruben M. Otxoa}
\affiliation{Hitachi Cambridge Laboratory, J. J. Thomson Avenue, Cambridge, CB3 0HE, United Kingdom.}

\author{Josu Etxezarreta Martinez}
\email{jetxezarreta@unav.es}
\affiliation{Department of Basic Sciences, Tecnun - University of Navarra, 20018 San Sebastian, Spain.}

\begin{abstract}
Deciphering complex gene-gene interactions remains challenging in transcriptomics as traditional methods often miss higher-order and nonlinear dependencies. This study introduces a quantum-inspired framework leveraging tensor networks (TNs) to optimally map expression data into a lower dimensional representation preserving biological locality. Using Quantum Mutual Information (QMI), a nonparametric measure natural for tensor networks, we quantify gene dependencies and establish statistical significance via permutation testing. This constructs robust interaction networks where the edges reflect biologically meaningful relationships that are resilient to random chance. The approach effectively distinguishes true regulatory patterns from experimental noise and biological stochasticity. To test the proposed method, we recover a gene regulatory network consisted of six pathway genes from single-cell RNA sequencing data comprising over $28.000$ lymphoblastoid cells. Furthermore, we unveil several triadic regulatory mechanisms. By merging quantum physics inspired techniques with computational biology, our method provides novel insights into gene regulation, with applications in disease mechanisms and precision medicine.
\end{abstract}
\maketitle

\section{Introduction}
Computational analysis, and more precisely computational biology, has revolutionized the ability to understand complex biological processes \cite{Markowetz2017}. This branch of science studies biological systems, focusing on data analysis, mathematical modeling, and computational simulations. Furthermore, from the beginning of the XXIst century,  advances in high-performance technologies and digitization have led to an exponential growth in biological data \cite{Altaf2014}. 

A key milestone in this process was the \textit{Human Genome Project (HGP)}, completed in 2003, which marked the beginning of mass genomics by generating the first partial sequencing of the human genome \cite{Collins2003}. In 2005, the emergence of \textit{next-generation sequencing (NGS)} technologies drastically reduced the costs and time associated with genomic analysis, in turn increasing the amount of data generated \cite{Mardis2008}.

Simultaneously, the concept of \textit{"omics"} emerged to describe the systematic study of sets of molecules, such as genes (genomics), messenger RNA transcripts (transcriptomics), proteins (proteomics) and metabolites (metabolomics). In transcriptomics, technologies such as RNA-Seq make it possible to measure the amount and sequence of messenger RNA in an organism, tissue, or cell at a specific time \cite{Morin2008}. These tools not only extract meaningful information, but also enable the creation of predictive models to explore complex biological phenomena beyond traditional experimental capabilities. In this context, inference of functional relationships between genes has emerged as a key area in molecular biology \cite{wang2009}.

Under this framework, inference of \textit{gene regulatory networks (GRN)} has become an important area in molecular biology, providing a formal representation of gene-gene interactions using graph representations, $\mathrm{G(V,E)}$, where nodes, $\mathrm{V}$, correspond to genes while edges, $\mathrm{E}$, represent regulatory interactions measured by transcription factors. These networks are fundamental to understand molecular mechanisms, including gene functions and their regulation in cellular processes \cite{Friedman2000}. However, GRN reconstruction remains challenging because experimental data (e.g., gene expression) only provides indirect evidence of regulatory interactions—such as correlations or perturbation responses—rather than direct observations of causality. Computational methods must therefore contend with these ambiguities, alongside technical noise and biological complexity \cite{Margolin2006, faith2007, haury2012}.

 In order to infer GRNs from gene expression data, different methods have been developed. These methods use statistical models to detect dependencies between gene expression profiles and establish possible regulatory relationships between them. Common approaches include correlation networks \cite{zhang2005}, information-theoretic scores \cite{butte1999mutual, mousavian2016}, regression-based models \cite{Salleh2017, Kim2009}, Gaussian graphical models \cite{schafer2005}, Bayesian and Boolean networks \cite{Lahdesmaki2006, Ilya2002, Friedman2000} or dynamic models \cite{oates2012, morrissey2010}. It should be noted that each method has its own assumptions and therefore specific limitations that restrict its applicability to more complex biological systems \cite{Chen2018, Pratapa2020, Diaz2022}. For example, many traditional methodologies assume linear or simple relationships between variables, which is  not always true, and more importantly, none of these conventional methods fully exploit the simultaneous and inter-regulatory connections between all genes. As a result, new ways for GRN inference are being proposed, including using quantum computers to do so \cite{roman2023}.

Tensor Networks (TNs) are compact mathematical representations designed to model high-dimensional systems with complex correlation structures \cite{cirac2009, schollwock2011}. Originally, at the turn of the century, those were developed to address problems in quantum many-body systems \cite{orus2019} and fluids \cite{gourianov2025}, allowing efficient representation of complex  quantum states in high-dimensional systems. Recently, these techniques have found applications in machine learning and data science \cite{sengupta2022, wang2023}, due to their striking ability to handle large and highly correlated data with reduced computational complexity. Mathematically, TNs take advantage of decompositions such as \textit{singular value decomposition (SVD)} to capture correlations between system components, allowing less relevant dimensions to be efficiently truncated. For example, Matrix Product State (MPS) and Tree Tensor Network (TTN) models are two common TN structures, with applications in quantum many-body systems.

In this work, we propose a Tensor Network (TN)-based framework to infer Gene Regulatory Networks (GRNs) from single-cell transcriptomic data, overcoming the limitations of pairwise interaction models by capturing higher-order dependencies among genes. Unlike traditional approaches restricted to binary correlations, our TN formulation naturally encodes multi-way regulatory relationships through its multilinear structure, preserving biological context while enabling efficient computation. By representing gene interactions as interconnected low-rank tensors, our method maintains the intrinsic high-dimensional structure of expression data, revealing complex regulatory motifs. The following sections discuss the theoretical foundations of this approach, detail the methodology, and present experimental results that highlight the ability of TNs to reveal intricate regulatory mechanisms that may be overlooked by classical techniques. 
As a result, our TN-based GRN inference method is a flexible tool that can capture gene regulation relationships in an accurate and computationally tractable manner.
\section{Quantum Computation Theory}

Quantum computation leverages quantum mechanics to represent and manipulate information. Unlike classical bits, which can be either $0$ or $1$, quantum bits (\textit{qubits}) can exist in a superposition of both states
within a two-dimensional Hilbert space. The state of a single qubit is written as:
\begin{equation*} \label{eq:single_qubit}
    |\psi \rangle = \alpha |0\rangle + \beta |1\rangle, \quad \text{with } |\alpha|^2 + |\beta|^2 = 1,
\end{equation*}
where $\alpha, \beta \in \mathbb{C}$ are complex probability amplitudes. In our gene analogy, $|0\rangle$ represents an inactive gene and $|1\rangle$ an active gene; the gene's quantum state reflects a probabilistic combination of both.

Upon measurement in the computational basis $\{|0\rangle, |1\rangle\}$, the system collapses to the state $|0\rangle$ with probability $|\alpha|^2$ or $|1\rangle$ with probability $|\beta|^2$.

When considering two qubits, their joint state is formed via the tensor product:
\begin{equation*} \label{eq:two_qubits}
\begin{split}
    |\psi\rangle &= |\psi_1\rangle \otimes |\psi_2\rangle \\
    &= \alpha_1\alpha_2|00\rangle + \alpha_1\beta_2|01\rangle + \beta_1\alpha_2|10\rangle + \beta_1\beta_2|11\rangle.
\end{split}
\end{equation*}
This product assumes that the qubits are independent. If they interact, the system may exhibit \textit{entanglement}, a uniquely quantum phenomenon where states cannot be factorized into individual components.

For a general system of $N$ qubits, the global state is:
\begin{equation*} \label{eq:Nqubits_state}
\begin{split}
    |\psi \rangle &= \sum_{i_1, i_2, \dots, i_N} c_{i_1 i_2 \dots i_N} |i_1\rangle \otimes |i_2\rangle \otimes \dots \otimes |i_N\rangle \\
    &= \sum_{i_1, i_2, \dots, i_N} c_{i_1 i_2 \dots i_N} |i_1 i_2 ... i_N \rangle,
\end{split}
\end{equation*}
where $c_{i_1 \dots i_N} \in \mathbb{C}$ and $\displaystyle\sum_{i_1, i_2, \dots, i_N} |c_{i_1 \dots i_N}|^2 = 1$ ensures normalization.

This formalism allows us to represent multiple genes within a unified computational framework, where their joint behavior can be described using concepts inspired by quantum systems. Such representations capture dependent gene activity—including synergistic regulation \cite{maeder2013}, epistasis, or other higher-order interactions—without requiring physical quantum effects. Instead, the approach uses mathematical tools from quantum theory to efficiently model complex dependencies that classical methods might overlook.


\section{Matrix Product State: an architecture for GRN inference}\label{mth:MPS}

MPSs are compact tensor network representations of quantum many-body wavefunction, especially suited for systems with limited entanglement \cite{perez2006, Cirac_2021, schollwock2011}. Instead of explicitly storing the full wavefunction tensor, which scales exponentially with the number of subsystems, MPS decomposes the global state into a sequence of low-rank tensors, drastically reducing computational cost.

Consider a quantum system of $N$ qubits (each with local dimension $d=2$), whose wavefunction can be expressed in the computational basis as:
\begin{equation*}
    |\psi\rangle = \sum_{i_1, \dots, i_N \in \{0,1\}} c_{i_1 \dots i_N} |i_1 \dots i_N\rangle.
\end{equation*}
The full set of coefficients $c_{i_1 \dots i_N}$ forms a rank-$N$ tensor, denoted $\psi_{i_1 i_2 \dots i_N}$, containing $2^N$ complex entries. For small $N$ (e.g., $N=2$), that is the matrix:
\begin{equation*}
    \psi_{i_1 i_2} =
    \begin{pmatrix}
        c_{00} & c_{01} \\
        c_{10} & c_{11}
    \end{pmatrix},
\end{equation*}
but for larger $N$, the tensor becomes infeasible to store or manipulate directly. MPS addresses this by factorizing the global tensor into a chain of smaller tensors:
\begin{equation*}
    \psi_{i_1 i_2 \dots i_N} = \sum_{\alpha_1, \dots, \alpha_{N-1}} A^{[1]}_{i_1, \alpha_1} A^{[2]}_{\alpha_1, i_2, \alpha_2} \dots A^{[N]}_{\alpha_{N-1}, i_N},
\end{equation*}
where each $A^{[k]}$ is a local tensor with a bounded dimension (called bond dimension), controlling how much entanglement can be encoded between subsystems.

This decomposition is especially effective for one-dimensional systems with area-law entanglement and forms the backbone of the Density Matrix Renormalization Group (DMRG) algorithm \cite{white2005}. In our biological context, this structure allows scalable representation and computation over high-dimensional gene expression states unveiling high-order interactions with quantum-inspired methods.

\begin{figure}[ht]
    \centering
    \includegraphics[width=0.4\linewidth]{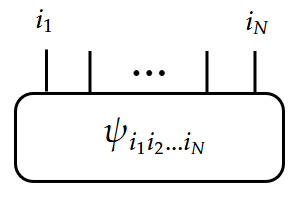}
    \caption{\small Rank-$N$ tensor visualized as a node with $N$ legs in the tensor network formalism. Each $i_j$ for any $j = 1,...,N$ plays the role of the gene `$g_{i}$'.}
    \label{fig:Ntensor_representation}
\end{figure}

To construct an MPS, we begin by reshaping the full wavefunction tensor $\psi_{i_1 i_2 \dots i_N}$ into a matrix that isolates the first subsystem:
\begin{equation*}
    \psi_{i_1 i_2 \dots i_N} \rightarrow \psi_{i_1 (i_2 \dots i_N)} \in \mathbb{C}^{d \times d^{N-1}}.
\end{equation*}
Applying a Singular Value Decomposition (SVD) to this matrix yields:
\begin{equation*}
    \psi_{i_1 (i_2 \dots i_N)} = \sum_{a_1 = 1}^{r_1} U_{i_1 a_1} S_{a_1} V^{\dagger}_{a_1 (i_2 \dots i_N)},
\end{equation*}
where $U$ and $V^{\dagger}$ are $d\times d$ and $d^{N-1}\times d^{N-1}$ unitary matrices, and $S$ is a diagonal matrix. The rank $r_1$ defines the first bond dimension. 
The first MPS tensor is given by $A^{i_1}_{a_1} = U_{i_1 a_1}$, while the remaining part is encoded in a reduced tensor:
\begin{equation*}
    \psi_{i_1 (i_2 \dots i_N)} = \sum_{a_1 = 1}^{r_1} A^{i_1}_{a_1} \psi_{a_1 (i_2 \dots i_N)}.
\end{equation*}

This process is repeated: the new tensor $\psi_{a_1 (i_2 \dots i_N)}$ is reshaped into a matrix of size $r_1 d \times d^{N-2}$, and another SVD is performed , yielding a new bond dimension $r_2$. Each step generates a new MPS  $A^{i_k}_{a_{k-1} a_k}$ with bond dimension $r_{k-1} \times r_k$.

After $N$ iterations, the full wavefunction is factorized as:
\begin{equation*}
    \psi_{i_1 i_2 \dots i_N} = \sum_{a_1,\dots,a_{N-1}} A^{i_1}_{a_1} A^{i_2}_{a_1 a_2} \dots A^{i_N}_{a_{N-1}},
\end{equation*}
or simply,
\begin{equation*}
    \psi_{i_1 i_2 \dots i_N} = A^{i_1} A^{i_2} \cdots A^{i_N},
\end{equation*}
where contractions over internal indices are implied.

\begin{figure}[ht]
    \centering
    \includegraphics[width=0.65\linewidth]{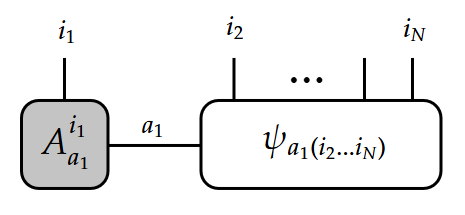}
    \caption{\small First step of MPS decomposition via SVD.}
    \label{fig:MPS_first_step}
\end{figure}

Each tensor $A^{i_k}_{a_{k-1} a_k}$ in the chain stores local information and correlations with neighboring subsystems. 
The size of the auxiliary indices  (bond dimensions) reflect the amount of entanglement captured at each truncation step.

In our context, gene expression states (binarized via Gaussian Mixture Models) \cite{mclachlan2019} are encoded as qubits ($d=2$), and MPS provides an efficient way to model their joint distribution. As described above, a MPS tensor network implies an ordering for the tensors. 
Thus, to preserve biological locality in this one-dimensional structure, we use a Hilbert space-filling curve to reorder genes, ensuring nearby genes in the 2D expression space remain adjacent in the 1D MPS chain. Hilbert space-filling curves have been used in the literature for more efficient representations of 2D quantum many-body systems into MPS tensor networks \cite{hilbertcurve}.

\begin{figure}[ht]
    \centering
    \includegraphics[width=0.65\linewidth]{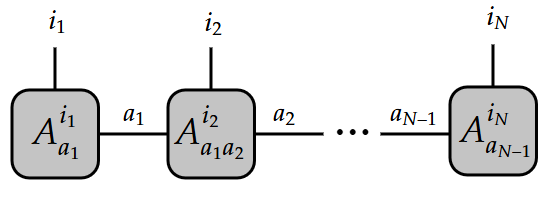}
    \caption{\small Final step: the full wavefunction as a product of $N$ local tensors.}
    \label{fig:MPS_last_step}
\end{figure}


\subsection{Gene expression binarization data}\label{mth:GMMs}

Gaussian Mixture Models (GMMs) are probabilistic models that are particularly effective for capturing heterogeneous or multimodal distributions, such as those often observed in gene expression data. In our context, they are used to model the distribution of expression levels for each gene across multiple patients, under the assumption that the expression follows a bimodal pattern: one mode corresponding to low (inactive) expression, and the other to high (active) expression.

Given a gene expression matrix \( X \in \mathbb{R}^{N \times m} \), where \( N \) is the number of genes and \( m \) is the number of patients, we aim to construct a binarized state matrix \( Z \in \{0,1\}^{N \times m} \). Each entry \( z_{ij} \) encodes the activity of gene \( i \) in patient \( j \), taking the value \( 1 \) if the gene is considered active, and \( 0 \) otherwise.

To achieve this, we fit a Gaussian Mixture Model with \( K = 2 \) components to the expression profile of each gene. The model assumes that each observed expression value arises from a mixture of two Gaussian distributions, corresponding to the two biological states of interest. Each component \( k \in \{1,2\} \) is characterized by the following unknown parameters:

\begin{itemize}
    \item \( \pi_k \): the mixture proportion (or prior probability) of component \( k \), representing how prevalent each state is in the population.
    \item \( \mu_k \): the mean expression level of component \( k \), indicating the typical activity level for that state.
    \item \( \sigma_k^2 \): the variance of component \( k \), accounting for biological variability and measurement noise.
\end{itemize}

These parameters define the marginal probability distribution of an observed expression value \( x \) as:
\begin{equation*} \label{eq:marginal_gmm}
\mathrm{P}(X_i = x) = \sum_{k=1}^{K} \pi_k \cdot \mathcal{N}(x; \mu_k, \sigma_k^2),
\end{equation*}
where $\mathcal{N}(x; \mu_k, \sigma_k^2)$ denotes the probability density function of a Gaussian with mean $\mu_k$ and variance $\sigma_k^2$. The weights satisfy $\pi_k > 0$ and $\sum_{k=1}^{K} \pi_k = 1$. Assuming that $X_1, \dots, X_n $ are independent and identically distributed (i.i.d.), the joint probability of the data is given by:
\begin{equation*} \label{eq:joint_gmm}
\mathrm{P}(X_1, \dots, X_n) = \prod_{i=1}^{n} \sum_{k=1}^{K} \pi_k \cdot \mathcal{N}(x_i; \mu_k, \sigma_k^2).
\end{equation*}

Collectively, these unknowns form the set of parameters $\theta = \{ \pi_k, \mu_k, \sigma_k^2 \}_{k=1}^{2}$, which is estimated from the data using the Expectation-Maximization (EM) algorithm. This iterative procedure consists of:

\begin{itemize}
    \item \textbf{E-step}: Computing the posterior probability that each expression value was generated by each component, based on the current parameter estimates.
    \item \textbf{M-step}: Updating the parameters $\theta$  to maximize the likelihood of the observed data, weighted by the posterior probabilities.
\end{itemize}

The algorithm repeats these steps until convergence, defined as a point where successive parameter estimates change negligibly. Once the GMM is fitted for each gene, we assign each gene expression value to the component (state) with the highest posterior probability. This binarization step enables us to represent gene expression data in a discrete latent space, while still preserving the probabilistic structure inferred from the continuous measurements.


\subsection{Hilbert Projection of Gene Expression}\label{mth:Hilbert}

To reduce the dimensionality for tensor network analysis, we first apply Principal Component Analysis (PCA) to the gene expression matrix \( X \in \mathbb{R}^{N \times m} \) or its binarized form \( Z \in \{0,1\}^{N \times m} \). 
Projecting onto the space formed by the top two principal components yields 2D coordinates \( (x_j, y_j) \) for each gene \( j \) . To preserve spatial locality in one dimension, we map these coordinates using the Hilbert space-filling curve \( H: (x, y) \rightarrow h \). Although not bijective, the Hilbert curve effectively preserves neighborhood structure better than other mappings \cite{Cataldi_2021}, making it ideal for structured representations. This is especially valuable in tensor network models, where locality improves computational efficiency \cite{mclachlan2019}. The resulting 1D representation \( h_j \) maintains key spatial relationships and it is well-suited for downstream analysis.


\subsection{Classical Data Translation to a State Vector Representation}

After binarization and projection onto a one-dimensional space via the Hilbert curve, each sample in the dataset is represented as a binary sequence $z_j \in \{0,1\}^{N}$, where $N $ is the number of genes ordered along the curve. This defines a state space $\mathcal{H}$  of dimension $2^N$, capturing all possible gene expression configurations. To construct a quantum representation, we compute the empirical probability $p_j$ of each unique sequence  $z_j$:
\begin{equation*}
    p_j = \frac{\#z_j}{m},
\end{equation*}
where $\#z_j$ is the frequency of  $z_j$ in the dataset and $m$ is the number of patients. Then, the quantum state vector isgiven by:
\begin{equation*}
    |\psi\rangle = \sum_{x_j \in \{0,1\}^{N}} \sqrt{p_j} \, |x_j\rangle,
\end{equation*}
where $|x_j\rangle = |i_1 i_2 \dots i_N\rangle$  represents a basis state in the $N$-qubit Hilbert space. The coefficients $\sqrt{p_j}$ ensure proper normalization. This encoding naturally aligns with the tensor network formalism, where quantum entanglement reflects statistical correlations between genes \cite{orus2019}. Using this representation, classical gene expression data can be efficiently mapped into a quantum-inspired computational framework for further analysis. In fact, we use this state vector to construct the MPS tensor network.


\subsection{Quantum mutual Information for gene regulation inference}\label{mth:QMI}
Once the data has been mapped to a quantum state and processed into a MPS tensor network, some measure must be taken over it to infer the regulation relationships. In quantum systems, mutual information extends to quantum mutual information (QMI), which quantifies both classical and quantum correlations between two subsystems $A$ and $B$. For a bipartite system with density matrix \( \rho_{AB} \), QMI is defined as:
\begin{equation} \label{eq:quantum_MI}
    \mathcal{I}(A, B) = S(\rho_A) + S(\rho_B) - S(\rho_{AB}),
\end{equation}
where \( S(\rho) = -\mathrm{Tr}(\rho \ln \rho) \) is the von Neumann entropy, and \( \rho_A \) and \( \rho_B \) are the reduced density matrices of $\rho_{AB}$ on subsystems $A$ and $B$, respectively. This QMI quantifies all correlations--both classical and quantum--between the subsystems, making it a natural choice for identifying potential regulatory relationships between genes. By computing $\mathcal{I}(A, B)$ for gene pairs, we can systematically detect significant gene-gene interactions within the biological network.

In the MPS framework, the reduced density matrices for QMI are computed by tracing out parts of the system. This is achieved through tensor network contractions to extract the marginal probability distributions, which correspond to the single-site reduced density matrix $\rho_l$, and the joint probability distributions from the two-site reduced density matrix $\rho_{l_1 l_2}.$ These contractions make use of the left and right environment in the MPS, enabling an efficient calculation of the entropies. Detailed derivations are provided in Appendix~\ref{app:marginal_joint}.

\subsection{Statistical Validation via Permutation Test}\label{mth:permutation}

To assess the statistical significance of the observed QMI values that quantify gene-gene dependencies, we apply a permutation test. This is performed by randomly permuting the gene labels across samples a number of times, thereby generating a null distribution of QMI values under the hypothesis of no structured dependency between the gene pair.

Unlike a standard two-tailed test, which evaluates extremeness in both directions, we compute both \textit{right-tailed} and \textit{left-tailed} $p$-values separately. This distinction is particularly informative in the context of mutual information-based measures such as QMI, which are inherently non-negative and often asymmetrically distributed.

For each gene pair $(i, j)$, we compute:

\begin{align}
    p_{\text{right}}^{(i,j)} &= \frac{\text{count}\left(\mathcal{I}_{\text{perm}}^{(i,j)} \geq \mathcal{I}_{\text{obs}}^{(i,j)}\right) + 1}{N_{\text{perm}} + 1}, \\
    p_{\text{left}}^{(i,j)}  &= \frac{\text{count}\left(\mathcal{I}_{\text{perm}}^{(i,j)} \leq \mathcal{I}_{\text{obs}}^{(i,j)}\right) + 1}{N_{\text{perm}} + 1},
\end{align}

where $\mathcal{I}_{\text{obs}}^{(i,j)}$ is the observed QMI for the gene pair $(i,j)$, $\{\mathcal{I}_{\text{perm}}^{(i,j)}\}$ are the permuted QMI values, and $N_{\text{perm}}$ is the number of permutations. The $+1$ correction prevents zero-valued $p$-values and ensures stability with low $N_{\text{perm}}$.

We report right- and left-tailed $p$-values separately, rather than combining them into a single two-tailed measure. This approach is motivated by the distinct biological interpretations of each tail. A significant right-tailed value indicates a gene pair with stronger than expected dependency \cite{marbach2012wisdom}. Conversely, a significant left-tailed value, indicates a pair with significantly weaker dependency than expected by chance.  This often reflects active biological mechanisms enforcing independence—such as mutual repression, compensatory regulation, or spatial segregation— \cite{canisius2016novel, ciriello2012mutual}.

\begin{figure*}[ht]
    \centering
    \includegraphics[width = 1\linewidth]{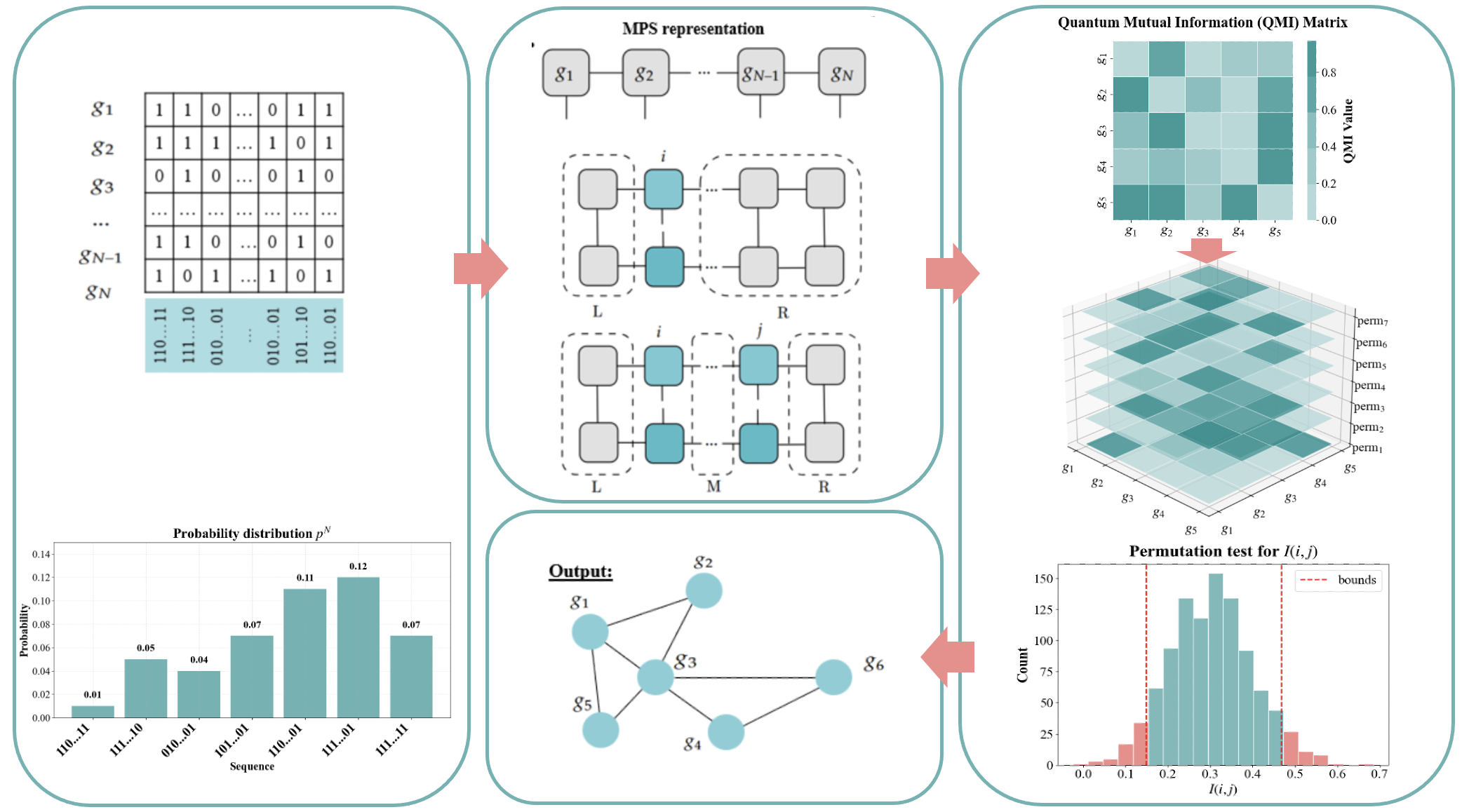}
    \caption{\small Schematic workflow of the Tensor Network based GRN inference method: (1) Gene $(g_i)$ expression binarization (GMM) and Hilbert curve ordering into a state vector $\ket{\psi}$, (2) MPS decomposition for entropy calculations are compression methods, (3) QMI matrix computation and permutation testing ($1000$ runs).}
    \label{fig:scheme_figure}
\end{figure*}

\section{Results}

NF-$\kappa$B (nuclear factor kappa-light-chain-enhancer of activated B cells) is a family of transcription factors central to the regulation of immune responses, cell survival, and differentiation. In B cells- which are a type of white blood cell-, NF-$\kappa$B dynamics—particularly the activity of its subunits c-Rel and RELA—play a pivotal role in the transition from proliferation to plasma cell differentiation. This transition is orchestrated through a tightly regulated gene regulatory circuit involving key transcriptional regulators such as IRF4, PAX5, PRDM1, and AICDA. Understanding the interplay among these genes is essential to decipher the regulatory logic driving B cell fate decisions.

In Fig.~\ref{fig:scheme_figure}, we show a diagrammatic representation of the workflow proposed to infer GRNs involving the techniques described above. In summary, the proposed method involves the following key computational steps: (1) data binarization and dimensionality reduction, (2) MPS representation and information extraction, and (3) statistical validation of inferred interactions.

To test the tensor network-based approach, we used it to reconstruct a gene regulatory network (GRN) consisted of six pathway genes (IRF4, REL, PAX5, RELA, PRDM1 and AICDA) from single-cell RNA sequencing (scRNA-seq) data comprising over 28.000 lymphoblastoid cells. The data has been accessed from the Gene Expression Omnibus (GEO) database \cite{GEO} with accession numbers: GSE126321 and GSE158275. 

In the following, we describe the processing of each of the steps required to infer the GRN embedded in said database and the obtained result. Furthermore, we employ our method to reveal certain triadic regulatory relationships observed in the literature, further showing the applicability of the tensor network-based method to recover complex correlations among genes.


\subsubsection{\noindent\textbf{Binarization and projection of gene expression data}}
Gene expression states were binarized (inactive = 0, active = 1) using Gaussian Mixture Models (GMMs), which automatically determined optimal activation thresholds for each gene while preserving biologically relevant expression patterns. 

We evaluated the bimodal structure by comparing GMM via the Bayesian Information Criterion (BIC) \cite{neath2012}, criterion for model selection among a finite set of models. For models of one and two parameters, the difference between them for each gene resulted in: $\Delta \mathrm{BIC_{12}} = 17512.0$ for AICDA,  ($\Delta \mathrm{BIC_{12}} = 9479.8$ for IRF4, $\Delta \mathrm{BIC_{12}} = 10587.5$, for PAX5, $\Delta \mathrm{BIC_{12}} = 13051.9$ for PRDM1, $\Delta \mathrm{BIC_{12}} = 9033.9$ for REL, and $\Delta \mathrm{BIC_{12}} = 10315.1$ for RELA. Further comparison between two- and three-component models revealed minimal improvement for more complex fits, e.g., $\Delta \mathrm{BIC_{23}} = 469.0$ for PAX5 and $\Delta \mathrm{BIC_{23}} = 92.8$ for RELA), supporting the two-component model as optimal. This aligns with the expected binary nature of gene activation while avoiding overfitting.

The binarized gene expression data was optimally ordered using a Hilbert space-filling curve \cite{hilbertcurve}, which is preserved $90.4 \%$ of the pairwise gene-gene expression correlations (Spearman $\rho = 0.904, \text{ $p$-value} < 10^{-4}$) between adjacent genes in the original high-dimensional space. This metric quantifies how well the 1D ordering maintains the rank correlation structure of neighboring genes. This projection mapped the genes into the order PRDM1, PAX5, AICDA, RELA, REL, IRF4. Crucially, the strong pairwise neighborhood  preservation indicates that the Hilbert curve ordering respects the original high-dimensional structure of the transcriptomic data, enabling faithful representation as a quantum state vector $\ket{\psi}$ for subsequent MPS decomposition and multi-gene interaction analysis.


\subsubsection{\noindent\textbf{Quantum mutual information matrix and statistical Validation}}
\begin{figure}[ht]
    \centering
    \includegraphics[width = 0.8\linewidth]{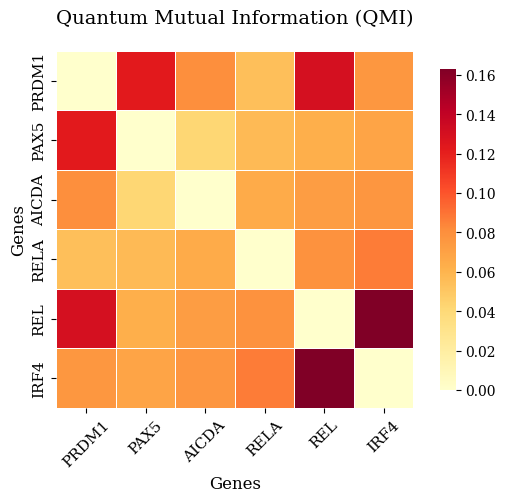}
    \caption{\small Quantum Mutual Information (QMI) matrix using Hilbert curve for GRN inference. }
    \label{fig:QMI_matrix}
\end{figure}

As described before, we measure the Quantum Mutual Information (QMI) between the tensors of the constructed MPS in order to infer the structure of the GRN in question. The resulting QMI matrix revealed the interaction landscape of the six NF-$\kappa$B genes, which we depict in Fig. \ref{fig:QMI_matrix}. 
\newline
In order to finish the GRN inference, we tested which of the obtained QMI values are statistically significant. In this way, we could determine which of the gene pairs present non-trivial QMI values indicating strong relationship among them. To do so, we realized one-tailed-left and-rigth permutation tests ($1,000$ iterations), constructing gene-specific null distributions to compute the empirical $p$-values shown in Tables ~\ref{tab:permutation_test_left_pvalues} and \ref{tab:permutation_test_rigth_pvalues}.

\begin{table}[ht]
\centering
\caption{\small Permutation test $p$-values of gene interactions - one-tailed-left}
\label{tab:permutation_test_left_pvalues}
\footnotesize
\begin{tabular}{ccccccc}
\toprule
 & PRDM1 & PAX5   & AICDA  & RELA   & REL    & IRF4 \\
PRDM1   & 1.0000 & 0.9990 & 0.9271 & 0.1099 & 0.9900 & 0.0869 \\
PAX5    & 0.9990 & 1.0000 & 0.0010 & 0.1938 & 0.0779 & 0.0090 \\
AICDA   & 0.9271 & 0.0010 & 1.0000 & 0.7752 & 0.5674 & 0.1898 \\ 
RELA    & 0.1099 & 0.1938 & 0.7752 & 1.0000 & 0.0820 & 0.7423 \\
REL     & 0.9990 & 0.0779 & 0.5674 & 0.9211 & 1.0000 & 0.2368 \\
IRF4    & 0.0869 & 0.0090 & 0.1898 & 0.7423 & 0.2368 & 1.0000 \\
\end{tabular}
\end{table}

\begin{table}[ht]
\centering
\caption{\small Permutation test $p$-values of gene interactions - one-tailed-rigth}
\label{tab:permutation_test_rigth_pvalues}
\footnotesize
\begin{tabular}{ccccccc}
\toprule
 & PRDM1 & PAX5   & AICDA  & RELA   & REL    & IRF4 \\
PRDM1   & 1.0000 & 0.0010 & 0.0729 & 0.8901 & 0.0010 & 0.9131 \\
PAX5    & 0.0001 & 1.0000 & 0.9990 & 0.8061 & 0.9221 & 0.9910 \\
AICDA   & 0.0729 & 0.9990 & 1.0000 & 0.2248 & 0.4326 & 0.8102 \\ 
RELA    & 0.8901 & 0.8061 & 0.2248 & 1.0000 & 0.0789 & 0.2577 \\
REL     & 0.0010 & 0.9221 & 0.4326 & 0.0789 & 1.0000 & 0.7632 \\
IRF4    & 0.92131 & 0.9910 & 0.8102 & 0.2577 & 0.7632 & 1.0000 \\
\end{tabular}
\end{table}

\begin{figure}[ht]
    \centering
    \includegraphics[width = 1\linewidth]{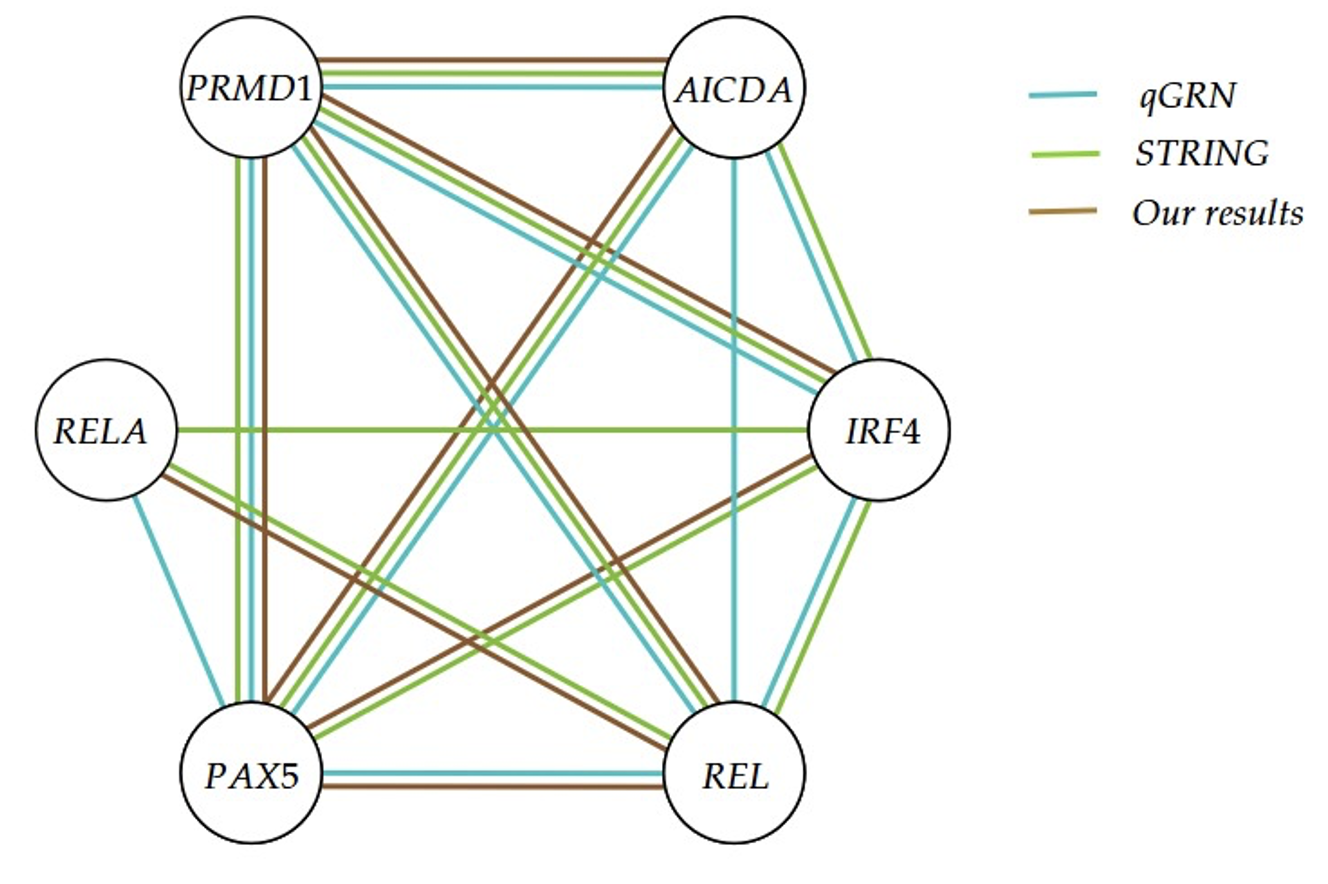}
    \caption{ \small Comparison of gene regulatory networks (GRNs) in B cells. (a) blue: quantum-based network (qGRN) obtained from \cite{roman2023}, (b) green: Interactions predicted by STRING, and (c) brown: Results of our analysis using TN. Edges represent significant regulatory interactions ($p < 0.1$).}
    \label{fig:Graph_GRN}
\end{figure}

In terms of results, the proposed method using tensor networks has identified different gene-gene interactions in lymphoblastoid B cells (LCLs), which are primary human B lymphocytes transformed with the Epstein-Barr virus (EBV). This approach has confirmed relevant interactions through significant values (p < 0.1), many of which coincide with previous studies \cite{roy2019regulatory, sciammas2011}. Among them is the PAX5–AICDA relationship \cite{yadav2006, Gonda2003}, in which PAX5 induces the expression of AICDA in activated B cells. Likewise, the relationship between PRDM1 (Blimp-1) and PAX5 \cite{boi2015prdm1} is recovered, where PRDM1 directly represses PAX5 expression, thus allowing the transition from the B cell's own gene program to one associated with plasma differentiation. This mechanism represents negative feedback \cite{nera2006loss}, and in B-cell lymphomas it is common to observe PAX5 overexpression or PRDM1 functional loss.

Relationships such as PRDM1–REL and PAX5–REL \cite{roman2023}, previously described, were also detected. Furthermore, unlike Cai's study \cite{roman2023}, our approach has identified an additional interaction strongly supported both experimentally and computationally according to STRING, RELA-REL \cite{milanovic2017, grossmann2000anti}.

Another notable finding is the detection of the interaction between PAX5 and IRF4, not reported by Cai \cite{roman2023}, but validated in STRING and in previous studies \cite{nera2006pax5, maffei2023dynamic}. These studies have described that high levels of IRF4 repress PAX5 expression through the activation of PRDM1, which, as mentioned above, silences PAX5, thus establishing cross-regulation. This observation reinforces the need to analyze the PAX5–PRDM1–IRF4 triad in more detail, which will be addressed in the section on multigenic relationships.

The method also identifies the PRDM1–IRF4 interaction, consistent with previous evidence \cite{Sciammas2006, tunyaplin2004direct}, where IRF4 cooperates with Bcl6 in the regulation of AICDA. When IRF4 levels are low, AICDA is activated; however, with progressive accumulation of IRF4, PRDM1 expression is induced, leading to the repression of AICDA. Likewise, a direct interaction between PRDM1 and AICDA is observed \cite{nutt2011genetic}, where PRDM1 represses AICDA to stop antibody diversification. However, unlike Cai \cite{roman2023} and the information collected in STRING, our analysis does not detect a direct relationship between IRF4 and AICDA. This finding suggests that the interaction could be mediated by a cascade effect, in which IRF4 regulates PRDM1, which in turn represses AICDA.

\subsubsection{\noindent{\textbf{Multi-Gene Relationships}}}
While our primary analysis focuses on pairwise quantum mutual information (QMI), gene regulatory systems often involve higher-order dependencies. Triadic interactions can reveal synergistic or redundant control not captured by pairwise metrics.

While there is less documented information on these more complex relationships, our approach presents the flexibility for doing so and, thus, we compute triadic QMIs as,
\begin{equation}
\begin{split}
    I(i;j;k) &= S(i) + S(j) + S(k) - S(ij) - S(ik)\\
    &- S(jk) + S(ijk),
\end{split}
\end{equation}
where $S(\cdot)$ denotes the von Neumann entropy. Positive $I(i;j;k)$ indicates synergy and negative values suggest redundancy, both biologically significant.

In line with the above, we have taken an additional step in characterizing more complex relationships by analyzing gene triads. This approach allows for a more comprehensive evaluation of functional interactions between genes that may not be fully apparent when analyzing pairs in isolation. In this context, one of the most notable triads is that formed by PRDM1, PAX5, and IRF4, whose tripartite interaction was statistically significant (p = 0.0110). This finding is consistent with previously observed gene-gene interactions and suggests that IRF4 could act as an upstream regulator of both PAX5 and PRDM1. This hypothesis is reinforced by previous studies describing how elevated levels of IRF4 induce the expression of PRDM1, which in turn represses PAX5, thus establishing a negative feedback loop between these genes. The joint identification of this triad suggests a possible coordinated regulatory module.

On the other hand, the triad composed of PRDM1, AICDA, and IRF4 did not show statistical significance in our analysis (p = 0.7030), which could be interpreted in light of the regulatory mechanisms described above. Although IRF4 is known to participate in the regulation of AICDA, this effect may be mediated indirectly, through the induction of PRDM1, which directly represses AICDA. Therefore, the absence of a statistically significant signal in this specific triad may support the hypothesis that the IRF4 → AICDA interaction is not direct, but part of a hierarchical cascade in which PRDM1 acts as an intermediary. This observation highlights the importance of considering multigenic structures rather than simple connections, as functional dependencies may depend on intermediate pathways not captured by traditional bivariate correlations or measures.

\section{Discussion}

In this article, we have introduced an MPS based approach to determine gene regulatory networks. 
The proposed method involves data binarization to map biological data to a state vector data, and Hilbert curve to select optimal gene ordering for constructing an MPS. Then QMI is measured to determine statistically significant regulatory relationships. We have shown that the method is succesful in recovering a GRN consisted of six pathway genes (IRF4, REL, PAX5, RELA, PRDM1 and AICDA) from single-cell RNA sequencing (scRNA-seq) data comprising over 28.000 lymphoblastoid cells.

Our tensor network approach shows significant advantages over other techniques in the literature, such as variational quantum algorithms (VQAs) for gene regulatory network (GRN) inference, particularly in the NISQ era. First, the VQA based qGRN pipeline proposed in \cite{roman2023} requires all-to-all qubit connectivity with circuit depth increasing as $\mathcal{O}\left(\binom{n}{2}\right)$, with $n$ being the number of genes. Such architectural constraints makes the approach challenging for most qubit platforms in which connectivity is usually restricted to nearest neighbors. Additionally, VQAs are highly susceptible to noise and limited coherence times (typically $ T_1 \ /T_2 \sim 100 \mu s$ \cite{kandala2019,toninid}) in current quantum hardware, which can degrade performance \cite{preskill2018, wang2021}. This is especially relevant for circuits with increasing depth complexity as the one in \cite{roman2023}. In fact, recent studies show that even moderate noise levels can exponentially suppress gradients in VQAs, a phenomenon exacerbated in NISQ devices \cite{wang2021, murali2019, Dalton2024}. In contrast, our classical tensor network implementation avoids these hardware limitations while retaining quantum-inspired efficiency.

Finally, VQAs are known to suffer from the so-called "barren plateau" problem, where gradients vanish exponentially with system size - even in the absence of noise- making optimization extremely challenging  \cite{mcclean2018, cerezo2021, holmes2022}. This issue is particularly relevant for the approach proposed in \cite{roman2023}, as their cost function is based on the Kullback-Leibler (KL) divergence, which requires computing a probability distribution over an exponentially large configuration space at each optimization step. In contrast, the MPS-based method proposed here operates directly on the data-driven probability distribution, avoiding the need for such costly and unstable optimization. This results in a more scalable and reliable framework for gene regulatory network (GRN) inference at mid-system sizes. Moreover, by relying on classical inference, our method allows for straightforward statistical significance testing, which would be highly non-trivial—and computationally expensive—in the quantum variational context.

Regarding the computational complexity of the approach, MPS scales polynomially $(\mathrm{O}(Nr^{3}))$ where $N$ are the genes with bond dimension $r$, making it computationally feasible for moderate-sized gene sets. 
However, in general, the bond dimension increases exponentially with the system size and, thus, exact MPS representation becomes difficult for large networks. To address this, adaptive truncation \cite{schollwock2011} methods can be employed to retain only the most significant singular values, balancing accuracy and efficiency. In addition, this approach is forward-compatible with future quantum hardware. Techniques such as tensor-to-qubit encodings via disentanglers \cite{MPScircuit} could enable hybrid quantum-classical implementations, ensuring scalability as quantum technologies mature.

Furthermore, we showed how tensor networks allow for scalable computation of triadic QMI across large gene sets, capturing these higher-order motifs. In future work, we will incorporate partial information decomposition (PID) \cite{williams2010nonnegative} to separate unique, redundant, and synergistic components. This could provide deeper mechanistic insights, particularly when integrated with single-cell multi-omics data. More complicated gene regulatory mechanisms may also be unveiled by using these techniques.

All in all, quantum-inspired frameworks offer promising tools to dissect multi-gene regulation in complex biological systems, with applications in decoding lineage programs and identifying network vulnerabilities in disease.

\section{Acknowledgements}
We thank other members of the Quantum Information Lab at Tecnun and the Hitachi-Cambridge Laboratory for their support and many useful discussions. We also thank members of Centro de Investigación Médica Aplicada (CIMA) of University of Navarra for comments and recommendations.

This work was supported by the Spanish Ministry of Economy and Competitiveness through the MADDIE project (Grant No. PID2022-137099NBC44); by the Diputación Foral de Gipuzkoa through the “Biased quantum error mitigation and applications of quantum computing to gene regulatory networks” project (2024-QUAN-000020); and by the Ministry for Digital Transformation and of Civil Service of the Spanish Government through the QUANTUM ENIA project call - QUANTUM SPAIN project, and by the European Union through the Recovery, Transformation and Resilience Plan - NextGenerationEU within the framework of the Digital Spain 2026 Agenda. We also acknowledge support from the Basque Quantum (BasQ) strategy. 


\appendix
\section{Marginal and Joint Probabilities in MPS} \label{app:marginal_joint}

\subsection{Marginal Probability Computation}
The marginal probability of a single subsystem in a MPS is obtained by tracing out all other subsystems. Given an MPS representation of a quantum state:
\begin{equation}
    |\psi\rangle = \sum_{\{i_k\}} A^{i_1} A^{i_2} \dots A^{i_N} |i_1 i_2 \dots i_N\rangle,
\end{equation}
the reduced density matrix of the \( l \)-th site, \( \rho_l \), is computed as:
\begin{equation}
    \rho_l = \mathrm{Tr}_{\neq l} \left( |\psi\rangle \langle \psi| \right).
\end{equation}
In the MPS formalism, this is achieved by contracting the left and right environments of the tensor \( A^{i_l} \), which encode all contributions from sites before and after \( l \). The left environment \( L \) and right environment \( R \) are defined as:
\begin{equation}
    L = \sum_{a_{l-1}, a'_{l-1}} \left( \prod_{k=1}^{l-1} A^{i_k}_{a_{k-1}a_k} \otimes \overline{A^{i_k}_{a'_{k-1}a'_k}} \right),
\end{equation}
\begin{equation}
    R = \sum_{a_l, a'_l} \left( \prod_{k=l+1}^{N} A^{i_k}_{a_{k-1}a_k} \otimes \overline{A^{i_k}_{a'_{k-1}a'_k}} \right).
\end{equation}
The resulting reduced density matrix is given by:
\begin{equation}
    \rho_l = \sum_{i_l, i'_l} L \cdot A^{i_l}_{a_{l-1}a_l} \cdot \overline{A^{i'_l}_{a'_{l-1}a'_l}} \cdot R.
\end{equation}
This contraction effectively traces out all subsystems except the \( l \)-th one, resulting in a \( d \times d \) matrix representing the probability distribution of states at site \( l \).

\subsection{Joint Probability Computation}
The joint probability of two subsystems \( l_1 \) and \( l_2 \) is obtained by computing the reduced density matrix \( \rho_{l_1 l_2} \), which traces out all other sites:
\begin{equation}
    \rho_{l_1 l_2} = \mathrm{Tr}_{\neq l_1, l_2} \left( |\psi\rangle \langle \psi| \right).
\end{equation}
This involves contracting three environments in the MPS representation:
\begin{itemize}
    \item The \textbf{left environment} \( L \), which accumulates contributions from sites before \( l_1 \).
    \item The \textbf{middle environment} \( M \), which contains the tensors \( A^{i_{l_1}} \) and \( A^{i_{l_2}} \) along with their indices.
    \item The \textbf{right environment} \( R \), which encodes contributions from sites after \( l_2 \).
\end{itemize}
The joint density matrix is computed as:
\begin{equation}
    \begin{split}
        \rho_{l_1 l_2} &= \sum_{i_{l_1}, i'_{l_1}, i_{l_2}, i'_{l_2}} L \cdot A^{i_{l_1}}_{a_{l_1-1}a_{l_1}} \cdot M \cdot A^{i_{l_2}}_{a_{l_2-1}a_{l_2}} \\ &\cdot \overline{A^{i'_{l_1}}_{a'_{l_1-1}a'_{l_1}}} \cdot \overline{A^{i'_{l_2}}_{a'_{l_2-1}a'_{l_2}}} \cdot R.
    \end{split}
\end{equation}
This results in a \( d^2 \times d^2 \) matrix that describes the joint probability distribution of subsystems \( l_1 \) and \( l_2 \).




\bibliographystyle{IEEEtran}
\bibliography{bibliography}


\end{document}